**The virtual drum circle: polyrhythmic music interactions in extended reality**

*Bavo Van Kerrebroeck, Kristel Crombé, Stéphanie Wilain, Marc Leman and Pieter-Jan Maes*

Department of Art, Music, and Theatre Sciences, IPEM, Ghent University, Ghent, Belgium

**Abstract**

Emerging technologies in the domain of extended reality offer rich, new possibilities for the study and practice of joint music performance. Apart from the technological challenges, bringing music players together in extended reality raises important questions on their performance and embodied coordination. In this study, we designed an extended reality platform to assess a remote, bidirectional polyrhythmic interaction between two players, mediated in real time by their three-dimensional embodied avatars and a shared, virtual "drum circle". We leveraged a multi-layered analysis framework to assess their performance quality, embodied coregulation and first-person interaction experience, using statistical techniques for timeseries analysis and mixed-effect regression and focusing on contrasts of visual coupling (not seeing / seeing as avatars / seeing as real) and auditory context (metronome / music). Results reveal that an auditory context with music improved the performance output as measured by a prediction error, increased movement energy and levels of experienced agency. Visual coupling impacted experiential qualities and induced prosocial effects with increased levels of partner realism resulting in increased levels of shared agency and self-other merging. Embodied coregulation between players was impacted by auditory context and visual coupling, suggesting prediction-based compensatory mechanisms to deal with the novelty, difficulty, and expressivity in the musical interaction. This study contributes to the understanding of music performance in extended reality by using a methodological approach to demonstrate how coregulation between players is impacted by visual coupling and auditory context and provides a basis and future directions for further action-oriented research.

## The virtual drum circle: polyrhythmic music interactions in extended reality

Human joint action can be considered one of nature's most remarkable features, occurring across a myriad of activities in which people coordinate actions towards a shared goal. Ample research have studied joint actions (Sebanz & Knoblich, 2021), deepening insights into the underlying behavioural and neuronal control principles (Koban et al., 2019; Richardson et al., 2015). Action-oriented research has been rewarding for investigations into cognition, stemming forth from the notion that cognition subserves action and is grounded in sensorimotor skills (Engel et al., 2016). For instance, research has shown how coordination through joint action is able to cause prosocial effects such as participatory sense-making (De Jaegher & Di Paolo, 2007) and feelings of shared control or joint agency (Bolt & Loehr, 2017; Loehr, 2022). They are also able to modulate and merge self-other representations (Heggli et al., 2019; Tarr et al., 2014) due to co-activation of action perception networks (Overy & Molnar-Szakacs, 2009). In the last decades, the study of human joint action has been accelerated by the integration of various new technologies, such as motion capture devices, neuroimaging techniques for measuring (inter)brain activity, and more lately, displays and interfaces from the domain of extended reality. More widely, extended reality has been integrated in various scientific disciplines as an innovative methodological tool.

Extended reality settings, encompassing virtual, mixed and augmented reality, have certain qualities that make them ideally suited for experimental research. For instance, they offer high degrees of experimental control while retaining acceptable and continuously improving levels of realism (Blascovich et al., 2002; Parsons et al., 2017; Van Kerrebroeck et al., 2021). In addition, using extended reality technology has created new fundamental questions (Metzinger, 2018) and offers flexible and new ways of presenting experimental stimuli such as body swapping (De Oliveira et al., 2016), the illusion of backwards time travel (Friedman et al., 2014) and conversing with your future self (Senel & Slater, 2020). It allows efficient data collection and eases the sampling for, and replication of, experimental work (Blascovich et al., 2002). Extended reality for the arts has enabled new multimodal experiences and non-linear narratives endowing visitors with agency and virtual embodiment (Baker, 2017). Given these traits, it is not surprising how extended reality finds its way into domains such as the social and cognitive (neuro)sciences (Parsons et al., 2017), philosophy (Metzinger, 2018) and the arts (Baker, 2017; Turchet et al., 2021).

Musical expression and interaction in the arts and the sciences have welcomed and experimented with extended reality early-on (Cipresso et al., 2018). Music as a research field is generous, in that it brings together many disciplines. It allows the study of human behaviour, cognition, and perception as well as a focus on engineering and computer science challenges in the musical practices of creating, composing or improvising, interpreting and listening. As such, music in extended reality has attracted a lot of attention (Turchet et al., 2021) and motivated others to outline detailed research agendas (Çamci & Hamilton, 2020). However, despite its wide and rapid spread, combinations of multi-user setups involving real-time, expressive interactions and low-latency networking are only recently gaining increased attention (Hamilton, 2019; Loveridge, 2020; Pai et al., 2020; Schlagowski et al., 2022; Turchet et al., 2022).

Music also represents an ideal practice in which joint-action dynamics can be investigated as it involves anticipatory and adaptive processes through fine-grained, bodily coregulation and expressive intentions across multiple timescales (Keller, 2014; Leman, 2007). External rhythms play an important role in these embodied dynamics as shown by the entrainment of dancers' body movements to the musical pulse (Burger et al., 2014; Miura et al., 2013; Miyata et al., 2017; Naveda & Leman, 2010) and the integration of the music's metrical structure within the listeners' body parts (Keller, 2014; Toiviainen et al., 2010). These rhythms can also facilitate interpersonal motor coupling with shared periodic movements, which have been shown capable of predicting coordination.

When musicians interact successfully, they exchange a continuous bidirectional information flow that allows effective coupling into an organic whole with characteristic traits (Demos et al., 2018; Walton et al., 2018). This coupling has been shown to positively correlate with self-rated "goodness" of performance (Chang et al., 2019) involving embodied dynamics that reflect musical structure and expression (Demos et al., 2018). Investigating joint-action in music performance and interaction could thus help to shed light on the sensorimotor, affective, and cognitive processes facilitating coordination. In addition to music's coregulatory function, it can enhance learning (Moore et al., 2017) and induce positive prosocial effects (Stupacher et al., 2017). For instance, synchronizing with music has been shown to lead to a sense of connectedness between people (Demos et al., 2012) with (shared) intentions playing an important role (Goupil et al., 2021). Due to this social and interactive nature of music, music has been described as an embodied language stressing the sense of joint agency induced by the motor actions evoked by sound (Dell'Anna et al., 2021).

When people coordinate, they must rely on informational channels that allow coupling that is neither too rigid nor too loose, allowing exploration and exploitation of emerging dynamics (Kelso, 2009; Warren, 2006). Visual coupling has been shown to increase synchrony between singers (D'Amario et al., 2018; Palmer et al., 2019), people in rocking chairs (Demos et al., 2012), listeners (Dotov et al., 2021; Vuoskoski et al., 2016), pianists (Kawase, 2014), dancers (Chauvigné et al., 2019; De Bruyn et al., 2008) and to reduce variability and individual differences in coordinating duos (Miyata et al., 2017). Visual and auditory coupling can both facilitate this coordinating function with their relative influence potentially depending on context (Chauvigné et al., 2019; Demos et al., 2012; Desmet et al., 2009; Dotov et al., 2021; Keller, 2014). Alleviating or adding one sensory channel could lead to compensatory mechanisms (Keller, 2014) and enhanced responses (Chang et al., 2017; Dotov et al., 2021). Due to co-representation of contributions and performance, even the belief that people are acting in another room can influence coordination dynamics (Atmaca et al., 2011; Milward & Sebanz, 2016).

While joint-action research and investigations into dyadic coregulation has made a lot of progress (Sebanz & Knoblich, 2021), a significant limitation of much of this research is that it focuses on unintentional or controlled tapping tasks (Chauvigné et al., 2019; Konvalinka et al., 2010). As music is an intrinsically social activity (Brown & Knox, 2017), there is a need for investigations that appropriately balance ecological validity and experimental control (D'Ausilio et al., 2015; Dotov et al., 2021). Extended reality technologies promise to fill this methodological gap as well as offer new pathways for experimental research (Kothgassner & Felnhofer, 2020; Metzinger, 2018). They promise to offer social presence (Van Kerrebroeck et al., 2021), virtual embodiment (Kilteni et al., 2012), perspective taking (Herrera et al., 2018), unique affordances (Berthaut, 2020) and prosocial attitudes towards virtual characters (Tarr et al., 2018). However, care must be taken in the design of virtual settings such as for the requirement to animate virtual characters with human-like motion data (de Borst & de Gelder, 2015; Pan & Hamilton, 2018). As extended realities technologies are still improving on ecological validity, it remains unclear whether these new technologies can mediate the rich and fine-grained expressions and intentions inherent to musical interactions (Keller, 2014; Leman, 2007).

This study aims to investigate the potential of extended reality technologies for experimental, musicological research. It operationalizes this aim by investigating to what extent visual coupling in the form of a virtual partner influences embodied coregulatory patterns in a rhythmic, dyadic task in extended reality. Since music has been shown to influence coordination and increase prosocial effects in a controlled setting (Demos et al., 2012; Stupacher et al., 2017), this study also compares the impact of music on coordination and experiential dynamics to the impact of a metronome. This paper starts with a description of the experimental design including the experimental task and protocol. Next, the materials and methods section describes the extended reality system developed to perform this research together with the quantitative methods used to perform data analysis. This is followed by a section with experimental results presented using a framework describing interpersonal dynamics in a performative, embodied coregulation and experiential layer (Van Kerrebroeck et al., 2021). Results are interpreted in the discussion section using the embodied music cognition (Leman, 2007) and coordination dynamics (Kelso, 2009; Warren, 2006) frameworks together with suggestions for future work.

**Research question**

The main research question of this study was to what extent visual coupling influences interpersonal coordination dynamics of two people engaged in a drumming task (See Figure 1). More specifically, the goal was to see to what extent visual coupling and the auditory context could facilitate a successful performance, effective embodied coregulation and positive prosocial effects. Visual coupling was varied using the visibility of the partner across not-seeing, seeing-as-avatar and seeing-as-real levels (partner realism). The auditory context was varied by accompanying the performance either by a metronome or a polyrhythmic backing track (musical background). See Figure 2 for an overview of the experimental design.

A first hypothesis was that the polyrhythmic backing track would help with anticipatory and adaptive sensorimotor processes of participants as it was more information rich than the metronome (Repp, 2006). Participants would thus show a more precise performance of smaller onset asynchronies and smaller variability for the conditions with the polyrhythmic backing track. As the guiding, visual stimulus disappeared during successful coordination periods (see Task below), the second hypothesis was that participants had to compensate for this loss of support by relying more on their bodily coordination (D'Amario et al., 2018; Miyata et al., 2017; Palmer et al., 2019). With increased partner realism, there should thus be an increase in shared periodic movements at the common pulse and, potentially, periodic movements at the polyrhythmic tempi across participants. A final hypothesis was that increased partner realism and music as background would be motivational for players and increase prosocial effects leading to higher scores of (shared) agency, self-other merging, and flow (Bolt & Loehr, 2017; Demos et al., 2012; Stupacher et al., 2017).

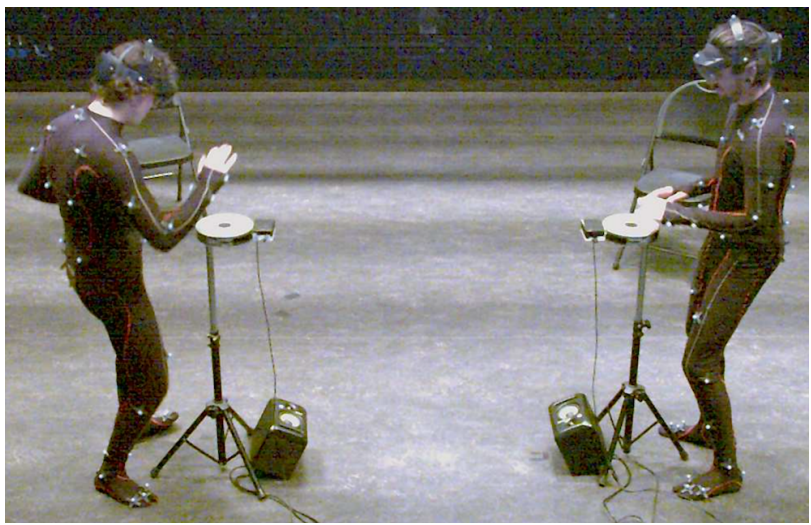

*Figure 1: Participants in the seeing-as-real condition*

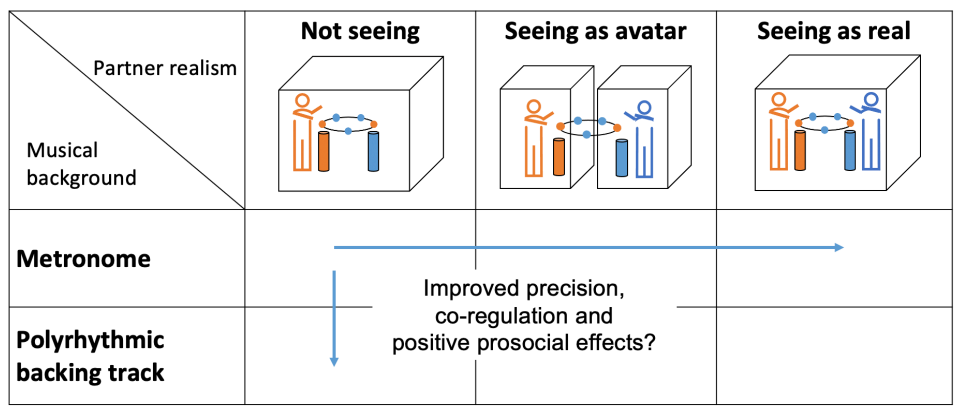

*Figure 2: Experimental conditions with the "partner realism" and "musical background" factors and hypotheses*

**Method**

*Participants*

16 dyads (13 women, 19 men, [M, SD]$_{age}$ = [30.2, 8.4] years, range: 18-48 years) with varying levels of musical expertise (musical training scores from the Musical Sophistication Index (MSI) (Müllensiefen et al., 2014), [M, SD] = [4.1, 1.9], range = 1-6.57) were recruited to participate in the experiment. 15 participants with MSI musical training scores higher than 3 out of 6 played percussion and piano instruments (8 percussionists, 7 pianists). Participants in each dyad were recruited together and knew each other beforehand to make them feel more at ease interacting with the avatar of their musical partner. There were no restrictions on gender, age, and relationship. The experimental protocol was reviewed and approved by the ethical commission of Ghent University, Belgium. Covid pandemic restrictions in effect were social distancing to 1.5 meter and the wearing of face masks when moving between labs.

*Materials and apparatus*

The experiment took place in the Art and Science Interaction Lab (ASIL) of Ghent University, Belgium. The technical set-up for this experiment was extensive and an overview is shown in Figure 3. Motion capture and avatars were used to investigate the full-body interactions between players. A virtual drum circle, drum pads and questionnaires were used to investigate performative and experiential aspects. All are listed below. Machines used in the setup are listed in the appendix.

*Motion capture:* Participants wore full-body motion capture suits with 42 markers. Participants were tracked using independent Qualisys set-ups in two separate rooms (about 7 m distance from one another).

*Extended reality – embodied avatars:* Qualisys software was used to render real-time, human-controlled, avatar visualizations in a Unity server application. These avatars were then streamed wirelessly to two HoloLens 2 head-mounted displays for both participants using 5Ghz WiFi routers under the 802.11ac standard. See Figure 4 for a first-person view of the avatar visualisation.

*Extended reality - drum circle:* A virtual drum circle with rotating spheres was rendered in Unity and displayed in both participant's HoloLenses. Note onset times of participants were compared to the timings of spheres passing in front of participants using a Max for Live application. See Figure 4 for a first-person view of the rotating spheres and drum circle.

*Networking:* All machines ran on the same local network and were directly connected to a HP5406 networking switch with 16 10GBps and 64 1Gbps connections, all compliant with the CAT6 standard. Audio routing was done using multiple Focusrite RedNet audio-over-IP interfaces using the Dante protocol with buffer sizes set at 128 samples and sample rates at 48kHz. MIDI notes, drum circle transparency, and skeleton data were sent using the Open Sound Control (OSC) protocol via UDP for lowest latency. 3D rendering on the two Unity clients were synchronized via a Unity server application built using the Unity Photon networking framework. A Focusrite Rednet Nanosync device generated a clock signal at 120Hz to keep audio, visuals, and motion capture data synchronized at the different machines.

*Latencies:* Latency from drum pad to speaker in either room was 17±2ms (5 measurements). Latency from motion capture marker to visualization in a Hololens was 58±4ms (5 measurements). Latency from drum pad into Max for Live was less than 5ms. Latency from Max for Live into Unity using the OSC protocol was 1.5±0.5ms (5 measurements). Dante audio networking latency was set to 1ms.

*Drum pads:* Participants triggered sounds by tapping on drum pads that sent MIDI notes to Ableton software which rendered them in a percussive sound[1]. Drum pads were custom built and registered note onsets using a pressure sensor connected to a Teensy 3.2 microcontroller. Drum pads were approximately 17 cm in diameter and set at waist height.

*Questionnaires:* Self-reports of participants were taken using Microsoft Forms questionnaires upon arrival and after each trial. They were asked to indicate whether they had received percussion training in the past, the length and intensity of the dyad's relationship, evaluate their relationship with the self-other integration (SOI) questionnaire (Aron et al., 1992) and fill in the MSI questionnaire before the experiment. Agency, shared agency, SOI and absorption scores from the Flow questionnaire (Engeser & Rheinberg, 2008) were recorded after each trial. Scales are presented in the analysis framework section below.

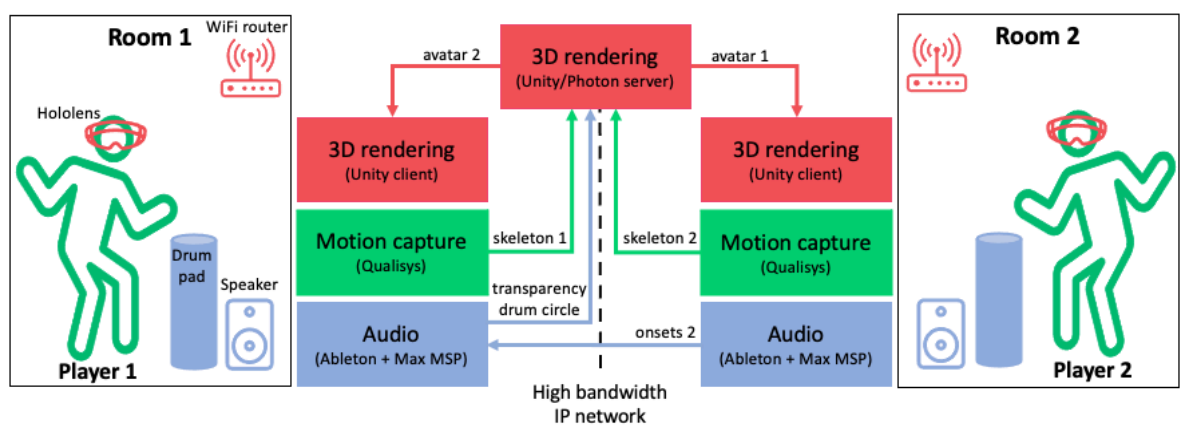

*Figure 3: The extended reality system for real-time, network, full-body interaction (red = 3D rendering, green = motion tracking, blue = audio)*

---

[1] See https://youtu.be/S7LjGwHRqyY for drum sounds of participants with polyrhythmic backing track

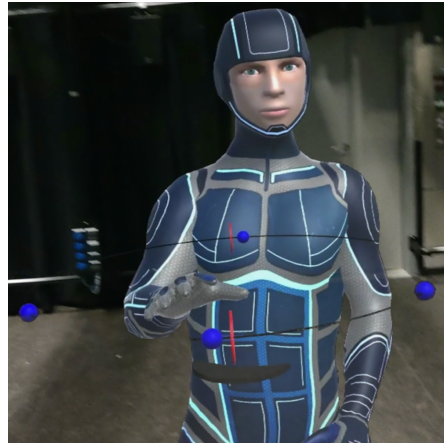

*Figure 4: View of the human-controlled avatar and the drum circle as rhythmic stimulus*

## Task

Dyads were instructed to tap a 2:3 polyrhythmic pattern together at a fixed tempo. Per dyad, one participant was instructed to tap the binary part of the polyrhythm (Inter Onset Interval (IOI) = 1059ms, further referred to as the 'binary task'), while the other was instructed to tap the ternary part (IOI = 706ms, further referred to as the 'ternary task'). Hence, the smallest unit (tactus) had a duration of 353ms, while the common pulse occurred every 2118ms.

The metronome consisted of a bell sound at pulse level that was also present in the polyrhythmic backing track. The backing track additionally consisted of the standard bell pattern (Agawu, 2006) with other percussive sounds playing on the tactus level[1]. Each condition lasted 242 seconds so participants could perform the polyrhythmic 2:3 pattern a maximum of 114 times.

An augmented reality visual stimulus in the shape of a drum circle was shown in every condition to assist with drumming (see Figure 4). The isochronous sequences for each participant were indicated as instructions by rotating virtual spheres on the drum circle and participants were allowed to improvise by skipping taps. When both participants successfully drummed the polyrhythmic pattern, meaning both players tapped within 62.5ms before or after their instructed tap, the stimulus increased in transparency in 5 incremental steps. When they made mistakes, the stimulus gradually returned.

The task was designed to be challenging enough to lead to alternating successful and unsuccessful periods of coordination in tapping the polyrhythmic pattern. The freedom to improvise was included to add aspects of playfulness and expressivity as well as to render the task more challenging for participants with musical experience. The disappearing instruction was designed to induce a form of reward and punishment according to reinforcement learning principles.

## Experimental design

Dyads performed polyrhythmic music interactions in six conditions (see Figure 2). Experimental factors were partner realism with 3 levels (not-seeing, seeing-as-avatar, seeing-as-real) and musical background with 2 levels (metronome, polyrhythmic backing track). We measured performative, embodied coregulatory and experiential outcome variables that are detailed in the Analysis framework section below. Conditions were randomized with the constraints of not-seeing and seeing-as-avatar as well as both the metronome and polyrhythmic back track conditions always performed together to decrease technical overhead in switching setups. An example order was SarMu-SarMe-SaaMe-SaaMu-NsMu-NsMe (labels are abbreviated from condition names, see Figure 2).

*Procedure*

The experiment lasted two hours. Upon arrival, health and safety rules were explained, and participants received a verbal explanation of the experiment, and gave their informed consent to participate in the experiment. They were then asked to change into motion capture suits and brought to one of two labs. After 45 minutes of motion tracking calibration and virtual skeleton building, both participants performed 2 minutes of their (binary or ternary) part of the task individually with a metronome and as instructed by the virtual drum circle. Next, while still being physically separated in different rooms, they were brought together virtually as human-controlled avatars to familiarize themselves with the real-time virtual animations (see Figure 4). After a brief exploration during which participants were encouraged to wave to each other and walk around, they performed another 2 minutes of the task together and with the polyrhythmic backing track as musical background. After these familiarizations, they were physically brought together to exchange some brief impressions and upon which they started the six conditions.

*Analysis framework*

The analysis framework is based on the earlier work of Van Kerrebroeck, Caruso, & Maes (Van Kerrebroeck et al., 2021). It consists of three interrelated layers containing performative, expressive and experiential dynamics between interacting players. The framework contains both qualitative self-reports and quantitative measures captured in real-time that are detailed below. A core feature of the framework is the comparison of interaction dynamics from a simulated, virtual context with the corresponding real-life context.

*Layer 1: Performance output:* The first layer was used to evaluate performance in relation to the goal prescribed by the musical task. The analysis was done using the BListener algorithm, a multivariate tracker of IOIs that uses Bayesian inference to predict timing constancy as a prediction error (Leman, 2021). It allows to deal with multiples of a basis IOI and is thus useful for dealing with player's improvisations that skip one or multiple tapping onsets in this study. Moreover, global features of timing constancy have shown to correlate with subjective estimates of performance quality and agency (Leman, 2021) and thus allow to link insights from this layer to the subjective layer below. Blistener initiates several trackers at the start of each performance that predict incoming IOI-data, which are continuously updated for each player and their instructions. Prediction errors of trackers are then summed over time leading to one global "prediction error" measure per trial for the joint performance of both players and for individual performances of player and (binary or ternary) instruction. BListener parameters, trackers for one trial and model parameters are presented in the appendix below. For more details about the BListener algorithm, the reader is referred to (Leman, 2021).

*Layer 2: Embodied coregulation:* This layer aimed at evaluating and comparing the embodied dynamics across conditions that players used to coordinate successfully. Movement data were used to compute postural position time-series and analysed using a Quantity of Motion (QoM) measure (Gonzalez-Sanchez et al., 2018) and wavelet coherence value (Grinsted et al., 2004). The instantaneous postural position was calculated by projecting the centre back marker of each player on the axis connecting the left and right foot markers. The QoM was calculated as the sum of all differences of consecutive samples of the postural position, that is, the first derivative of the position time series following the approach of Gonzalez-Sanchez et al. Wavelet coherence was calculated on concatenated postural sway timeseries of each player. Only parts of timeseries that had a transparent stimulus were kept to reduce the variability between players and focus the analysis on the good interaction bouts. A coherence value per trial was then obtained by summing respective coherence values in 0.2Hz frequency bands centred around the common pulse, the binary and ternary rhythms present in the instruction (common pulse: 2188ms or 0.472Hz, binary: 706ms or 1.416Hz, ternary: 1059ms or 0.944Hz).

*Layer 3: Subjective experience:* The goal of this layer was to evaluate subjectively experienced interaction qualities and link mental states, (expressive) intentions and meaning attributions to observations in other layers. Self-reports from questionnaires inquired about the length of participants' relationship, their SOI score, whether participants were trained in percussion, and the musical engagement and training factors from the MSI. Another questionnaire was taken after each trial and inquired about perceived quality of the performance, feelings of (shared) agency, SOI, and flow. Agency and shared agency scores were recorded using 7-point Likert scales as response to the question "To what extent did you feel agency/shared agency over the musical performance" (1 = Not at all, 4 = Partly, 7 = Very much).

*Statistical models:* Outcome measures from each layer were evaluated using linear mixed effect (LME) and cumulative link mixed (CLM) models. Variables and their interactions were progressively added to a null model based on whether they significantly improved the model fit. Variables are listed in Table 1 and final models in Table 2. Models in layer 1 and 2 included the dyad as a random effect while outcome measures from questionnaires in layer 3 included the individuals as a random effect. Models' intercepts corresponded to the player with the ternary task, not seeing the other binary player, in the first trial, playing with the metronome background and having lowest scores for self-reported scores. Outliers in the data were detected using 1.5 times the IQR as metric as well as after model fitting using residual quantile plots with two-sided outlier tests ($\alpha = .025$). The removed outliers are reported below.

*Data processing and analysis:* Timeseries trimming and alignment was done using Python. Wavelet coherence values reported below were calculated using Matlab. Other outcome measures such as the prediction error and quantity of motion were calculated using R and Matlab. Statistics were performed in R using the lme4 package (Bates et al., 2014) for the linear mixed effect models and the ordinal package (Christensen, 2015) for the cumulative link models. Tabular reports in the appendix were produced using the sjPlot package (Lüdecke, 2017) with analyses of variance, deviance and contrasts done using the Anova and the emmeans functions. Five trials from three dyads were excluded due to missing data in one trial and a misunderstanding of the task in the others.

| Type | Predictor | Levels |
|---|---|---|
| **Experimental conditions** | partner_realism | not-seeing, avatar, real |
|  | musical_background | metronome, music |
| **Experimental characteristics** | task | ternary, binary |
|  | trial_count | 1 to 6 |
| **Self-reports** | relation_frequency | 5-point Likert scale[2] |
|  | relation_length | number of months |
|  | percussion | 0 or 1 |

---

[2] 1=Never, 2=Less than once per month, 3=1-3 times per month, 4=1-3 times per week, 5=More than 3 times per week

|  | training | 1 to 7 |
|---|---|---|
|  | engagement | 1 to 7 |
|  | SOI | 1 to 7 |

Table 1: Experimental model terms

|  | **Outcome measures** | **Model** |
|---|---|---|
| **Layer 1** | prediction_error (individual performance) | 1 + (partner_realism + musical_background)*task + trial_count + training + relation_frequency + (1|dyad) |
|  | prediction_error (joint performance) | 1 + partner_realism + musical_background + trial_count + SOI + engagement + training + (1|dyad) |
| **Layer 2** | QoM | 1 + partner_realism + musical_background + trial_count + relation_length + SOI + (1|dyad) |
|  | coherence | 1 + partner_realism + musical_background + trial_count + training |
| **Layer 3** | agency | 1 + partner_realism + musical_background + trial_count + (1|individual) |
|  | shared_agency | 1 + partner_realism + trial_count + percussion + relation_length + (1|individual) |
|  | SOI | 1 + partner_realism + trial_count + training + (1|individual) |
|  | flow | 1 + partner_realism + musical_background + SOI + relation_frequency + relation_length + (1|individual) |

Table 2: LME and CLM models in each methodological layer

**Results**

This section reports the quantitative results from LME and CLM model fitting. Model statistics are presented for each methodological layer together with figures of outcome measure means, standard errors (SE) and model fits across levels of partner_realism and musical_background. A summary of all significant effects across layers is given in Table 3. Model parameters, confidence intervals and corresponding statistics are reported in the appendix.

*Layer 1: Performance output*

Two statistical analyses were done on the prediction errors obtained from the BListener algorithm: one focused on individual players with their instruction and one on the joint performance between players.

*Individual performance:* Eleven out of 182 data points (coming from 16 dyads performing in six conditions minus five excluded trials) were removed as outliers. Prediction error data were log-transformed to assure normally distributed model residuals. Exponentiated estimates and CIs are presented below. The model explained 63% of the variance with a majority portion explained by the fixed effects (conditional $R^2$ = 0.628, marginal $R^2$ = 0.442). There was a main effect of task ($\chi^2(1)$ = 44.2, $p < .001$) and musical background ($\chi^2(1)$ = 17.5, $p < .001$). Pair-wise comparisons revealed a significant difference between musical backgrounds for the binary ($t(146)$ = 4.172, $p < .001$) but not for the ternary (($t(146)$ = 1.794, $p = .075$) task. Task (ß = 1.78, CI = [1.43, 2.22], $t(151)$ = 5.22, $p < .001$) and musical training were significant predictors in the model (ß = 0.941, CI = [0.885, 1.00], $t(151)$ = -198, $p = .050$), while musical background was not ($p = .075$). Trial count ($\chi^2(5)$ = 21.8, $p < .001$) and relation frequency ($\chi^2(4)$ = 30.2, $p < .001$) showed significant effects with prediction errors linearly decreasing with trial count (ß = 0.799, CI = [0.700, 0.912], $t(151)$ = --3.35, $p = .001$) and linearly increasing with relation frequency (ß = 1.93, CI = [1.36, 2.73], t(151) = 3.73, *p* < *.001*). The 5th degree term of trial count (ß = 1.23, CI = [1.08, 1.40], t(151) = 3.07, *p* = *.003*) and the quadratic term of relation frequency (ß = 1.35, CI = [1.09, 1.68], t(151) = 2.78, *p* = *.006*) were significant as well.

*Joint performance:* Three out of 91 data points were removed as outliers. Prediction error data were log-transformed. The model explained 67% of the variance (conditional $R^2$ = 0.669, marginal $R^2$ = 0.375). Musical background ($\chi^2(1)$ = 6.60, *p* = *.010*) and engagement ($\chi^2(1)$ = 5.33, *p* = *.021*) were revealed as significant effects with the polyrhythmic backing track (ß = 0.801, CI = [0.674, 0.951], t(74) = -2.57, *p* = *.012*) and more engagement (ß = 0.769, CI = [0.612, 0.965], t(74) = -2.31, *p* = *.024*) decreasing the prediction error. Trial count was revealed as a significant main effect ($\chi^2(5)$ = 16.3, *p* = *.006*) with prediction errors decreasing linearly with trial count (ß = 0.762, CI = [0.617, 0.940], t(74) = -2.57, *p* = *.012*) together with a positive 5th degree term (ß = 1.35, CI = [1.10, 1.66], t(74) = 2.88, *p* = *.005*).

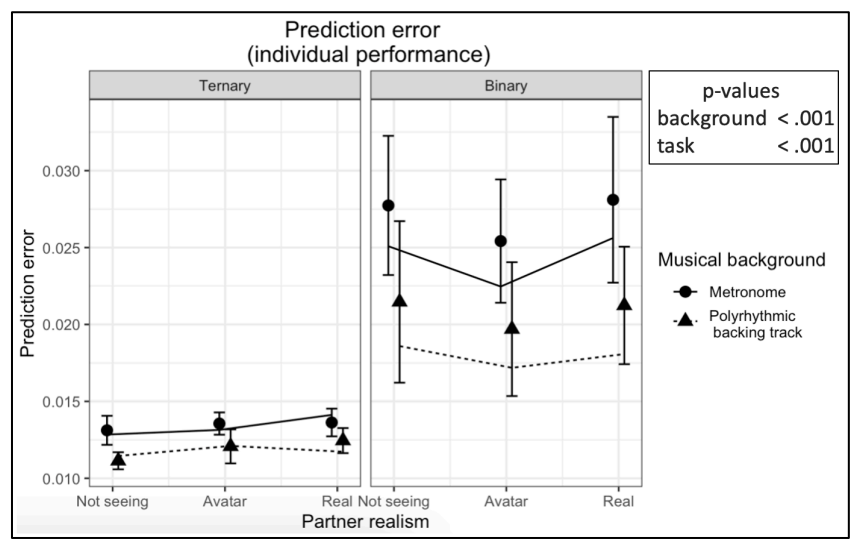

*Figure 5: Prediction error means, standard errors and LME model fits for the individual performance across partner realism and musical background*

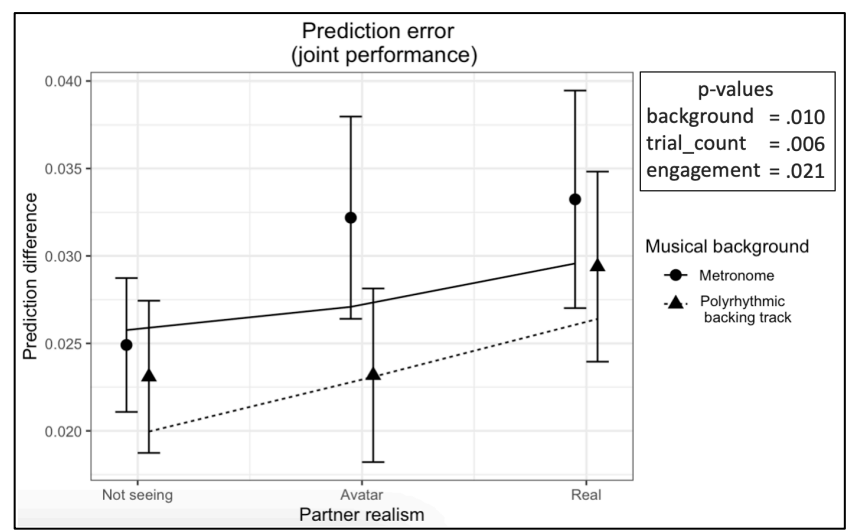

*Figure 6: Prediction error means, standard errors and LME model fits for the joint performance across partner realism and musical background*

*Layer 2: Embodied coregulation*

In this layer, one analysis was performed on the QoM of players individually and one on the wavelet coherence value of their joint performance. For the latter, only bandpower for the ternary rhythm was retained as mean bandpower for the common pulse (M = 0.19, SE = 0.003) and binary rhythm (M = 0.21, SE = 0.004) were significantly lower ($p < .001$) than ternary bandpower (M = 0.28, SE = 0.006). In addition, the binary bandpower did not significantly differ from power in bands at four (M = 0.21, SE = 0.003, $p = .907$) and five (M = 0.22, SE = 0.003, $p = .360$) times the tactus frequency.

*QoM:* The model explained 58% of the variance with 30% explained by the fixed effects (conditional $R^2$ = 0.58, marginal $R^2$ = 0.30). There were significant main effects for partner realism ($\chi^2(2) = 15.1, p < .001$), musical background ($\chi^2(1) = 40.2, p < .001$), trial count ($\chi^2(5) = 21.4, p < .001$), relation length ($\chi^2(1) = 9.7, p = .002$) and SOI ($\chi^2(5) = 21.0, p < .001$). Post-hoc tests revealed a significant difference between not seeing and seeing as avatar (t(154) = -0.12, SE = 0.038, $p = .003$) and not seeing and seeing as real (t(153) = 0.14, SE = 0.042, $p = .003$). The polyrhythmic backing track significantly increased the QoM (ß = 0.194, CI = [0.133, 0.254], t(165) = 6.34, $p < .001$) while the relation length decreased QoM (ß = -0.003, CI = [-0.005, -0.001], t(165) = -3.12, $p = .002$).

*Coherence:* Three out of 91 data points were removed as outliers. Wavelet coherence data were log-transformed to assure normally distributed model residuals. As including random effects did not improve model fit ($p = .5687$), they were not included in the model. The fitted linear model explained a significant and substantial portion of variance ($R^2$ = 0.298, F(9, 78) = 3.67, $p < .001$, adjusted $R^2$ = 0.217). Musical training (F(1, 78) = 6.12, $p = .016$) was a significant main effect, while musical background (F(1, 78) = 3.88, $p = .052$) and trial count (F(5, 78) = 2.27, $p = .055$) almost. Musical training increased the coherence (ß = 1.05, CI = [1.01, 1.09], t(78) = 2.47, $p = .016$) just like seeing as real (ß = 1.12, CI = [1.00, 1.24], t(78) = 2.07, $p = .042$), while a polyrhythmic backing track decreased it (ß = 0.924, CI = [0.853, 1.01], t(78) = -1.97, $p = .052$).

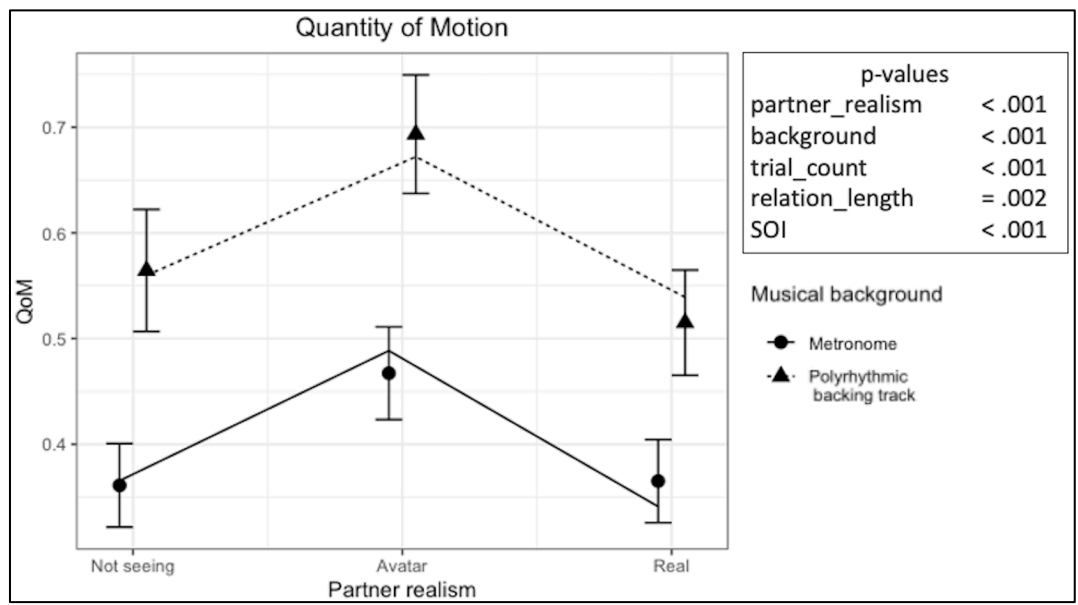

*Figure 7: Quantity of motion means, standard errors and LME model fits of players across partner realism and musical backgrounds*

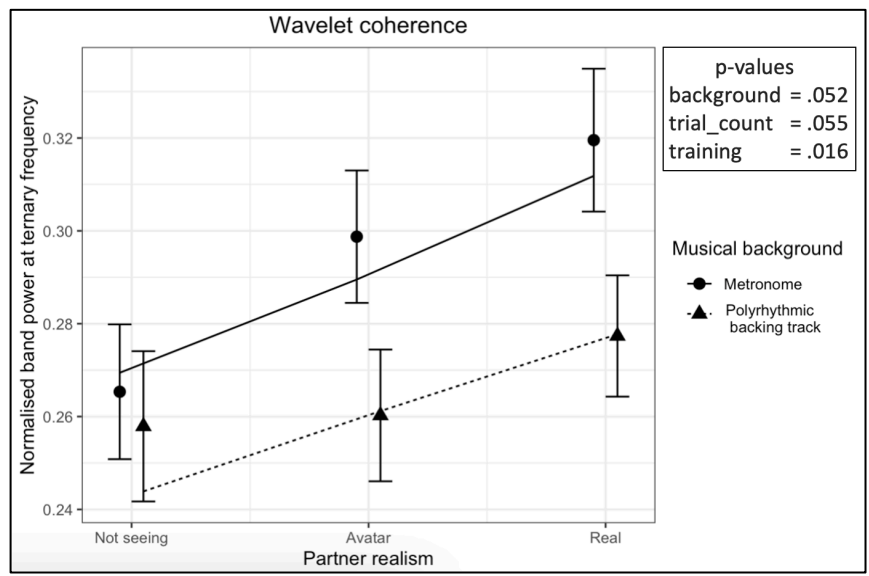

*Figure 8: Wavelet coherence in the ternary rhythm band means, standard errors and LM model fits across partner realism and musical backgrounds*

*Layer 3: Subjective experience*

This section presents the results from participant self-reports retrieved from questionnaires. Figures 9-11 below show model predictions of the agency, shared agency, flow, and SOI outcome measures.

*Agency:* The model explained 47% of the variance (conditional $R^2 = 0.469$, marginal $R^2 = 0.165$) and revealed a main effect of musical background ($\chi^2(1) = 12.0$, $p < .001$) and trial count ($\chi^2(5) = 34.0$, $p < .001$). Specifically, participants were 2.75 times more likely to report higher scores of agency when playing with the polyrhythmic backing track (odds-ratio (OR) = 2.75, CI = [1.54, 4.92], $p = .001$). Agency increased linearly (OR = 5.86, CI = [2.84, 12.1], $p < .001$) and decreased in the 5$^{th}$ degree (OR = 0.337, CI = [0.168, 0.680], $p = .002$) with trial count.

*Shared agency:* The model explained 39% of the variance (conditional $R^2 = 0.388$, marginal $R^2 = 0.228$). Main effects were revealed for partner realism ($\chi^2(2) = 16.6$, $p < .001$) and trial count ($\chi^2(5) = 26.4$, $p < .001$) with weaker effects revealed for percussion ($\chi^2(1) = 6.03$, $p = .014$) and relation length ($\chi^2(1) = 4.15$, $p = .042$). The seeing as avatar (OR = 3.12, CI = [1.55, 6.27], $p = .001$) and seeing as real (OR = 4.06, CI = [1.90, 8.71], $p < .001$) conditions were respectively 3.12 and 4.06 times more likely to result in higher shared agency scores compared to not seeing. Playing percussion also increased the odds of reporting higher feelings of shared agency (OR = 2.77, CI = [1.26, 6.07], $p = .011$) just like the trial count (OR = 4.83, CI = [2.37, 9.82], $p < .001$), while the relation length slightly decreased it (OR = 0.99, CI = [0.972, 0.999], $p = .036$).

*SOI:* The model explained 69% of the variance (conditional $R^2 = 0.694$, marginal $R^2 = 0.215$). Main effects were revealed for partner realism ($\chi^2(2) = 20.8$, $p < .001$), trial count ($\chi^2(5) = 48.9$, $p < .001$) and a weak effect for musical training ($\chi^2(1) = 3.61$, $p = .057$). Both seeing as avatar and seeing as real significantly increased the odds of reporting higher SOI compared to not seeing (avatar: OR = 2.32, CI = [2.1.15, 4.66], $p = .019$; real: OR = 6.49, CI = [2.83, 14.9], $p < .001$). Musical training was a significant, positive predictor in the model (OR = 1.53, CI = [1.02, 2.29], $p = .038$). Trial count had a strong, positive effect for its linear term (OR = 10.43, CI = [4.83, 22.5], $p < .001$) and weaker positive and negative effects respectively for its 4$^{th}$ (OR = 2.71, CI = [1.275, 5.77], $p = .010$) and 5$^{th}$ degree (OR = 0.476, CI = [0.241, 0.941, $p = .033$) terms.

*Flow:* Five out of 182 data points were removed as outliers. The model explained 54% of variance (conditional $R^2$ = 0.541, marginal $R^2$ = 0.297). Musical background ($\chi^2(1)$ = 8.46, *p = .004*), prior SOI score ($\chi^2(5)$ = 14.3, *p = .014*) and relation length ($\chi^2(1)$ = 6.02, *p = .014*) were significant main effects revealed by the model. The flow score increased with a polyrhythmic backing track (ß = 0.035, CI = [0.011, 0.059], *p = .004*) and with increased relation length (ß = 0.001, CI = [0.0002, 0.002], *p = .015*). Quadratic (ß = 0.109, CI = [0.033, 0.184], *p = .005*) and cubic (ß = 0.077, CI = [0.015, 0.138], *p = .016*) terms for the prior SOI score were significant predictors in the model.

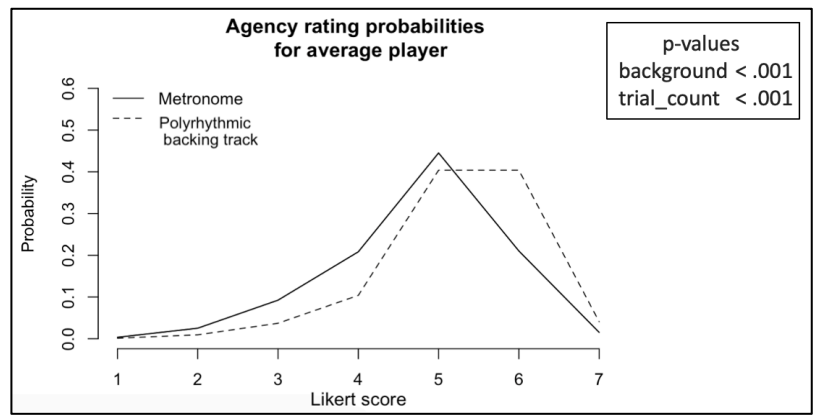

*Figure 9: Agency rating probabilities for an average player across musical backgrounds*

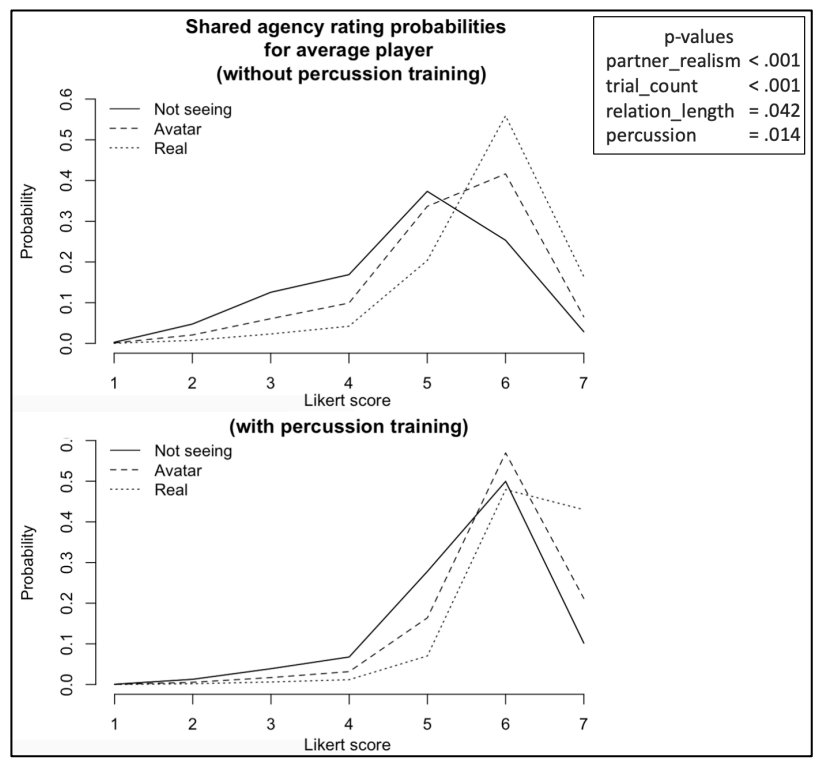

*Figure 10: Shared agency rating probabilities for an average player with and without percussion training across partner realism*

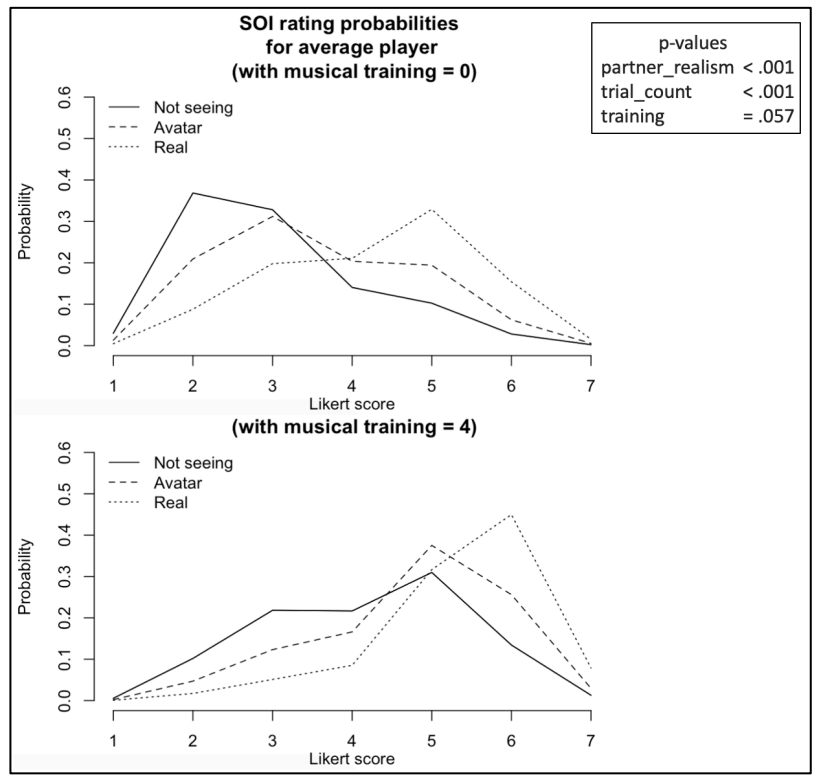

Figure 11: Self-other merging rating probabilities for an average player with a musical training equal to 0/7 or 4/7 across partner realism

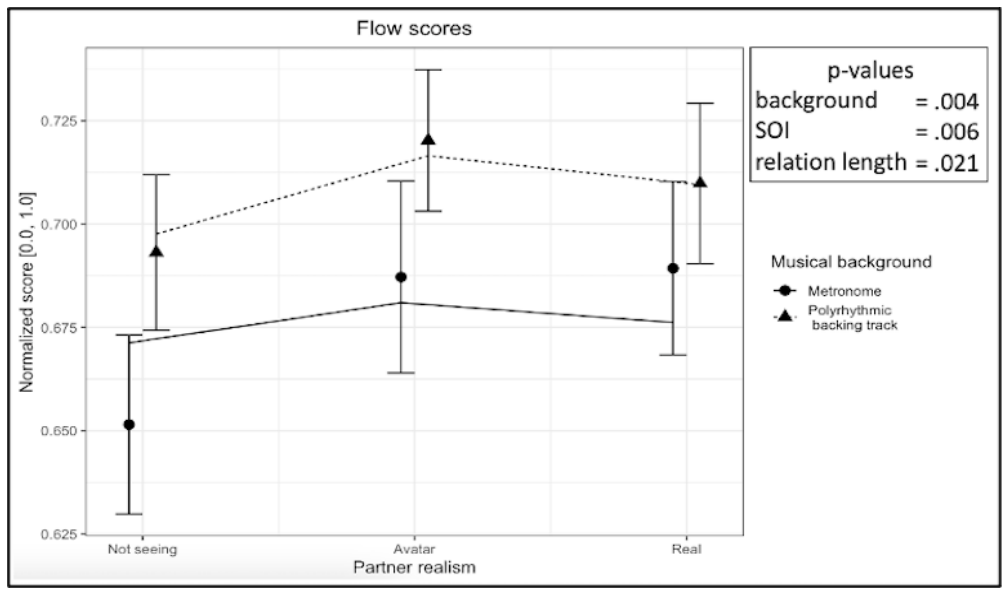

Figure 12: Flow scores, standard errors, and LME model fits for all players across partner realism and musical backgrounds

|  | **Outcome measure** | **Predictor** | **Significance** |
|---|---|---|---|
| **Layer 1** | prediction_error (individual performance) | musical_background | < .001 |
|  |  | task | < .001 |
|  | prediction_error (joint performance) | musical_background | .010 |
|  |  | trial_count | .006 |
|  |  | engagement | .021 |
| **Layer 2** | QoM | partner_realism | < .001 |
|  |  | musical_background | < .001 |
|  |  | trial_count | < .001 |
|  |  | relation_length | .002 |
|  |  | SOI | < .001 |
|  | coherence | musical_background | .052 |
|  |  | trial_count | .055 |
|  |  | musical training | .016 |
| **Layer 3** | agency | musical_background | < .001 |
|  |  | trial_count | < .001 |
|  | shared_agency | partner_realism | < .001 |
|  |  | trial_count | < .001 |
|  |  | relation_length | .042 |
|  |  | percussion | .014 |
|  | SOI | partner_realism | < .001 |
|  |  | trial_count | < .001 |
|  |  | musical training | .057 |
|  | flow | musical_background | .004 |
|  |  | SOI | .014 |
|  |  | relation_length | .014 |

*Table 3: Significant effects in each methodological layer*

**Discussion**

This experimental study aimed to investigate to what extent the coregulatory joint-action dynamics between two interacting, musical players are impacted when performing in an extended reality environment. It did this by operationalizing a methodological framework introduced in earlier research (Van Kerrebroeck et al., 2021) to analyse coregulatory dynamics across a performative, embodied and subjective layer while varying the players' visual coupling and auditory context. The next paragraphs will first discuss the results in each analysis layer, followed by this study's broader implications and directions for future research.

*Performance output:* Overall, players performing the binary rhythm had a larger prediction error and thus performed worse than the ternary rhythm. Given the fact that the binary rhythm's tempo was further away from a preferred "resonance" tempo (Van Noorden & Moelants, 1999) it is likely that this made it more difficult to correctly anticipate and adapt to upcoming pulses (Konvalinka et al., 2010; Repp & Su, 2013). Both players had comparable errors across levels of partner realism but were more accurate when playing with the polyrhythmic backing track and improved their playing over time. Learning effects had significant linear and 5$^{th}$ order terms with the latter illustrating the fact that trials were bundled (two musical backgrounds for each level of partner realism) due to the randomization constraints. Participants thus learned to perform the task better across as well as within levels of partner realism.

*Embodied coregulation:* Participants moved more energetically with the polyrhythmic backing track and when seeing the other as avatar. Given that the metronome background was more difficult to perform, the player might have compensated for this increased challenge using intensified bodily movements. Visual coupling might have served as a compensatory or communicative mechanism to deal with the lower information density in the metronome background. Analysis of coherence values indicated that participants moved less in synchrony when playing with the polyrhythmic backing track. Bodily coordination or coherence as synchrony between players' movements only showed convincing levels at the ternary rhythm. The absence of a clear coherence at the common pulse might reflect the difficulty of synchronizing with the slowest pulse (Van Noorden & Moelants, 1999) while the lack of coherence at the binary rhythm might be because it was more difficult to perform (Møller et al., 2021). Coherence at the ternary rhythm decreased over trials and was lower when playing with a polyrhythmic backing track. While the hypothesis was that players would coregulate their movements more when playing in a richer auditory context and when getting more familiar with each other and the task, this decreased coherence and the improvement of prediction errors over time suggest that each player increasingly embodied their binary or ternary rhythm individually (Burger et al., 2014; Toiviainen et al., 2010). While players had a shared goal (create the polyrhythmic pattern) with a shared pulse, this goal might have been too complex to achieve, motivating players to focus on their individual parts first. Further research could investigate whether this decrease in coherence also holds for expert musicians as musical training had a positive effect on coherence levels.

*Subjective experience:* Visual coupling of players led to higher feelings of shared agency and SOI. These effects held for both seeing as avatar as for seeing as real with a more outspoken effect for the latter. Agency and flow increased when playing with a polyrhythmic backing track. It appeared that seeing the other modulated player's perceived sense of shared control possibly because partner's actions became more predictable (Bolt & Loehr, 2017). Music, as an information and expressively rich mediator, on its turn showed more capable of modulating player's perceived sense of self control possibly through the increased movement related to the richer musical background (Dell'Anna et al., 2020). An interesting role was played by musical expertise in which more musical training and training in percussion led to a higher sense of SOI and shared agency respectively. These findings align with the prosocial effects of music found in other research (Stupacher et al., 2017) and might suggest the lesser role played by visual coupling for musicians to induce shared agency and SOI as compared to novices. As active investigations into the different forms of agency are underrepresented in virtual settings (Loehr, 2022), this study demonstrated an effective way to disentangle their relations and provided first steps towards leveraging the potential of extended reality for research into agency.

Circling back to the hypotheses in this study, the results indicate how a richer musical background led to the expected increase in performative precision between players' individual and joint performance. Given that the polyrhythmic backing track included beats at all levels (tactus, binary and ternary), the increased information density offered a rhythmic template on which to minimize asynchronies (Repp, 2006) and might have improved motor timing and stability through an entrainment effect (Carrer et al., 2022; Rose et al., 2021) and stronger groove (Leow et al., 2014). Regarding the second hypothesis, it was shown how increased partner realism and a richer musical background led to an increase in positive prosocial effects. These findings correspond to other research highlighting the social power of music (Stupacher et al., 2017; Tarr et al., 2014) and of synchronising with virtual humans (Hale & Hamilton, 2016; Tarr et al., 2018). It thus seems that the performative and experiential aspects of the dyadic interactions did not deteriorate in extended reality. Regarding the final hypothesis, different levels of partner realism did not have a significant impact on movement coherence between players while players did move more energetically when seeing each other as avatar and playing with a richer musical background. Novelty of the extended reality medium might have led to exaggerated movements to compensate for expressive limitations (Chang et al., 2017) as well as enhanced social contagion (Dotov et al., 2021).

Several limitations were present in this study. While extended reality has been praised for its capability of replication (Blascovich et al., 2002; Parsons et al., 2017), the high amount of expertise required to deal with the technological complexity and managing of different multimodal data streams makes replication more difficult. Furthermore, while audio latencies in the current system were within the 8 to 25ms region required for natural interaction (Chafe et al., 2010), visual stimuli followed auditory stimuli by approximately 40ms, a situation not encountered in real life. While this auditory-visual latency lies within the 30-50ms temporal integration window (Schwartz & Savariaux, 2014; Vroomen & Keetels, 2010) and approaches the 75ms maximal round-trip latency of earlier high-end networked music systems (Drioli et al., 2013), it does extend beyond the acceptable delay of 25ms for drumming synchrony reported in (Petrini et al., 2009). This lower bound should be taken with care though, given the latter study's small sample population (8 participants) and comparable perceived synchrony values for a delay of 67ms or less. Given the complex nature of multisensory integration (Harrar et al., 2017), with people compensating for audiovisual delays over time (Powers et al., 2009; Vroomen & Keetels, 2010) and musical expertise affecting temporal windows of integration (Petrini et al., 2009), future work should further quantify the impact of network latencies on sensorimotor coordination, not in the least considering the high emotional and attentional demands in musical contexts and the increased networked interactions we have today. Finally, while this study demonstrated an approach to perform empirical extended reality research in musical interactions, other limitations of this study are the relatively small sample size and a sample population containing both novices and musicians.

While players were allowed to improvise to induce more play in the musical interactions, this aspect was not analysed in this study. Future work could make improvisation and the role of visual coupling a research question on itself (Eerola et al., 2018) or include other aspects such as expressivity and underlying (shared) intentions (Goupil et al., 2021) in the layered analysis framework. While the polyrhythmic task was ideally suited for improvisation and playfulness, a synchronisation/continuation task such as in (Schultz & Palmer, 2019) might be an interesting alternative as it would more directly relate movement synchrony or coherence with task performance. While we opted to perform a cross-sectional study to attract a sample population with diverse backgrounds and characteristics, future work could opt to perform a longitudinal study (with novices or experts) to avoid large variability between dyads and focus on the observed learning effects. Finally, while the findings here result from a dyadic interaction between two virtual humans or avatars, future work could introduce additional human or computer-controlled players to investigate group dynamics (Dotov et al., 2022).

The analysis presented here has shown how virtual humans can be used to test specific hypotheses in a controlled experimental paradigm. Given the flexible nature of extended reality stimuli, building an extended reality platform such as the one in this study offers a solid basis on which to continue further investigations. For example, one could relatively easily replace the human-controlled avatars by computer-controlled agents (Tarr et al., 2018; Van Kerrebroeck et al., 2021) to test and refine dynamic models of control. By moving the research closer to where the action is (Engel et al., 2016), a paradigm such as the one presented in this study promises to achieve a more detailed view on the dynamics underlying social interaction and cognition.

## Conclusion

Extended reality technologies have tremendous benefits to offer for scientific research as they allow us to test specific hypotheses in controlled, experimental scenarios in ever more realistic models of the world. However, a crucial need for music, and by extension joint-action or action-oriented, research is to assure that coregulation in human coordination does not break down in an extended reality context. This study evaluated this potential of experimental settings in extended reality by investigating to what extent visual coupling and an auditory context between musical players engaged in a joint-action polyrhythmic task influences performative, embodied coregulatory dynamics and induces positive prosocial effects. While an informationally richer auditory context proved to be most beneficial for lower-level sensorimotor performative dynamics and favoured individual over joint performance, visual coupling in the form of an avatar or as in real life positively impacted experiential qualities of the interaction and induced prosocial effects. Embodied coregulation through bodily coordination was found to play an important role as illustrated by significant effects of visual coupling and auditory context on movement energy. With the analysis presented in this study and its relevance for a broad range of research themes ranging from sensorimotor synchronisation, embodied coordination to experiential aspects in musical interactions, this study has demonstrated the value of state-of-the-art extended reality technologies in action-oriented research. A key point has been to show how the introduction of networked, (human-) controlled virtual humans capable of a realistic level of real-time expressive (bodily) interaction offer new experimental paradigms with more ecological validity while retaining high levels of experimental control.

## Acknowledgements

The present study was funded by Bijzonder Onderzoeksfond (BOF) from Ghent University, Belgium. The drum pads were built by Ivan Schepers.

## Declaration of interest

The authors have no conflict of interest to declare.

## Appendix

| 1 x rendering server | CPU: i7-5820K @ 3.3GHz, RAM: 64GB, GPU: NVIDIA GeForce GTX TITAN X, Storage: 1TB SSD Intel 750, OS: W10e |
|---|---|
| 2 x rendering client | CPU: Intel Xeon 6136 @ 3GHz, RAM: 128GB, GPU: NVIDIA GeForce RTX 2080 Ti, Storage: 512GB Samsung 970, OS: W10e |
| 2 x motion capture | CPU: Intel Xeon 4110 @ 2.1GHz, RAM: 64GB, GPU: NVIDIA GeForce RTX 2070, Storage: 2TB SSD Samsung 970, OS: W10e |
| 2 x audio | CPU: i7-8700K @ 3.7GHz, RAM: 64GB, GPU: NVIDIA GeForce GTX 1080, Storage: 512GB + 1TB SSD 970 PRO, OS: W10e |

*Table 4: computing machines used in the experimental setup*

| Parameter | Value |
|---|---|
| Metervalue | 3/2, 2, 3, 4, 6, 8, 9 |
| Tgvalue | 1000 * 60/170 |
| Gravity | 0.01 |
| Outscope | 0.05 |
| System noise | 1e-05 |
| Observation noise | 1e-03 |
| Sampling rate | 100 |
| Number of cycles | 10 |
| Number of blocks | 4 |

*Table 5: Blistener parameters*

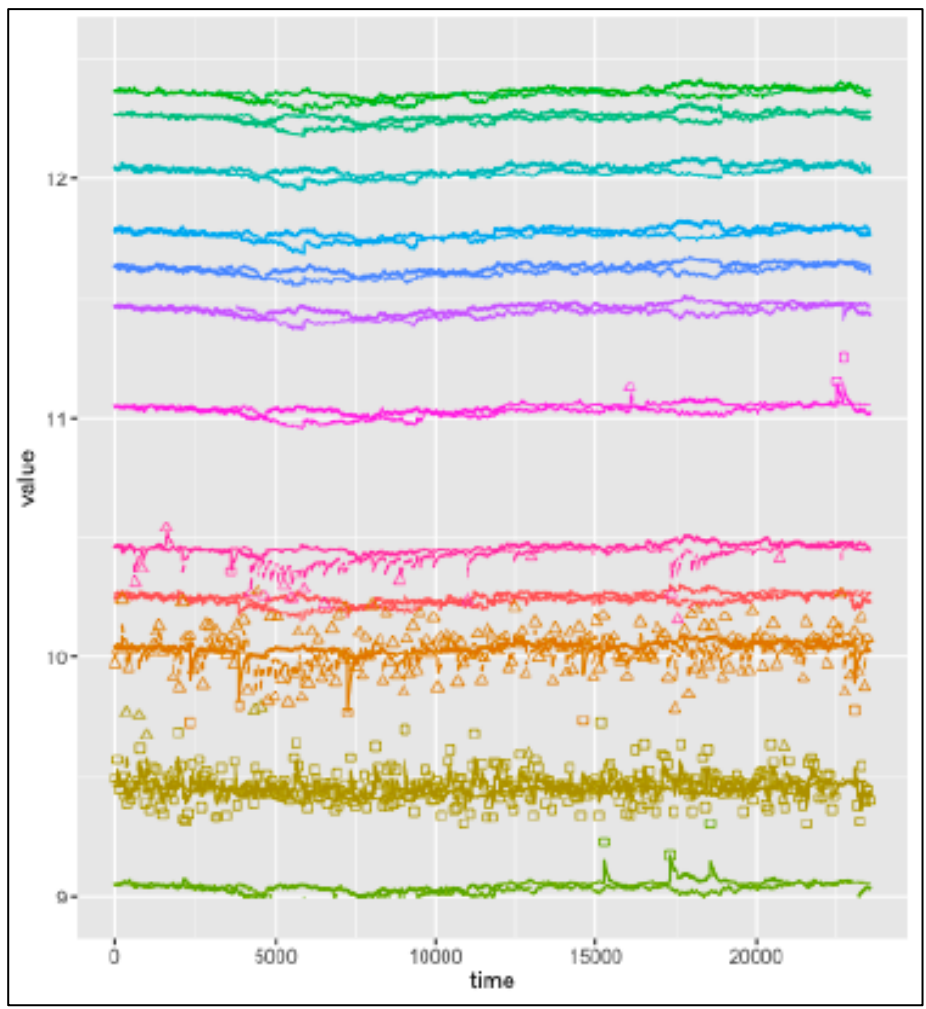

*Figure 13: IOI-predictions of 2 participants using 12 trackers per participant*

|  | Prediction error (individual performance) | | |
|---|---|---|---|
| *Predictors* | *Estimates* | *CI* | *p* |
| (Intercept) | 0.0194 | 0.0132 – 0.0284 | **<0.001** |
| partner realism [2] | 1.0550 | 0.8755 – 1.2713 | 0.572 |
| partner realism [3] | 1.1056 | 0.9091 – 1.3445 | 0.312 |
| musical background [1] | 0.8717 | 0.7493 – 1.0140 | 0.075 |
| task [Binary] | 1.7845 | 1.4334 – 2.2217 | **<0.001** |
| trial count [linear] | 0.7990 | 0.7000 – 0.9122 | **0.001** |
| trial count [quadratic] | 0.9928 | 0.8577 – 1.1492 | 0.922 |
| trial count [cubic] | 0.9478 | 0.8315 – 1.0805 | 0.420 |
| trial count [4th degree] | 1.0316 | 0.8969 – 1.1865 | 0.661 |
| trial count [5th degree] | 1.2265 | 1.0755 – 1.3987 | **0.003** |
| training | 0.9408 | 0.8853 – 0.9999 | **0.050** |
| relation frequency [linear] | 1.9290 | 1.3620 – 2.7321 | **<0.001** |
| relation frequency [quadratic] | 1.3547 | 1.0919 – 1.6808 | **0.006** |
| relation frequency [cubic] | 1.0412 | 0.8185 – 1.3244 | 0.741 |
| relation frequency [4th degree] | 0.8340 | 0.6576 – 1.0577 | 0.133 |
| partner realism [2] * task [Binary] | 0.8421 | 0.6485 – 1.0935 | 0.196 |
| partner realism [3] * task [Binary] | 0.9179 | 0.7058 – 1.1937 | 0.520 |
| musical background [1] * task [Binary] | 0.8316 | 0.6721 – 1.0290 | 0.089 |
| **Random Effects** | | | |
| $\sigma^2$ | 0.1234 | | |
| $\tau_{00\ dyad}$ | 0.0614 | | |
| ICC | 0.3323 | | |
| $N_{dyad}$ | 16 | | |
| Observations | 171 | | |
| Marginal $R^2$ / Conditional $R^2$ | 0.442 / 0.628 | | |

*Table 6: Overview of the LME model of the prediction error for the individual performances between players and (binary and ternary) task (estimates, confidence intervals, p-values, random effect and residual variances)*

|  | Prediction error (joint performance) | | |
|---|---|---|---|
| *Predictors* | *Estimates* | *CI* | *p* |
| (Intercept) | 0.0228 | 0.0176 – 0.0294 | **<0.001** |
| partner realism [2] | 1.0366 | 0.8387 – 1.2813 | 0.736 |
| partner realism [3] | 1.2673 | 0.9998 – 1.6064 | 0.050 |
| musical background [1] | 0.8007 | 0.6739 – 0.9513 | **0.012** |
| trial count [linear] | 0.7616 | 0.6168 – 0.9404 | **0.012** |
| trial count [quadratic] | 0.9414 | 0.7463 – 1.1876 | 0.606 |
| trial count [cubic] | 1.1065 | 0.8986 – 1.3626 | 0.336 |
| trial count [4th degree] | 0.9679 | 0.7724 – 1.2128 | 0.774 |
| trial count [5th degree] | 1.3482 | 1.0961 – 1.6583 | **0.005** |
| SOI | 0.8374 | 0.6690 – 1.0481 | 0.119 |
| engagement | 0.7686 | 0.6124 – 0.9645 | **0.024** |
| training | 0.8089 | 0.6518 – 1.0039 | 0.054 |
| **Random Effects** | | | |
| $\sigma^2$ | 0.1578 | | |
| $\tau_{00\ dyad}$ | 0.1398 | | |
| ICC | 0.4696 | | |
| $N_{dyad}$ | 16 | | |
| Observations | 88 | | |
| Marginal $R^2$ / Conditional $R^2$ | 0.375 / 0.669 | | |

*Table 7: Overview of the LME model of the prediction error for the joint performance between players (estimates, confidence intervals, p-values, random effect and residual variances)*

| Predictors | Quantity of motion | | |
|---|---|---|---|
| | Estimates | CI | p |
| (Intercept) | 0.4832 | 0.3536 – 0.6129 | **<0.001** |
| partner realism [2] | 0.1248 | 0.0501 – 0.1996 | **0.001** |
| partner realism [3] | -0.0141 | -0.0963 – 0.0682 | 0.736 |
| musical background [2] | 0.1936 | 0.1334 – 0.2539 | **<0.001** |
| trial count [linear] | 0.1544 | 0.0807 – 0.2281 | **<0.001** |
| trial count [quadratic] | 0.0042 | -0.0778 – 0.0861 | 0.920 |
| trial count [cubic] | 0.0032 | -0.0692 – 0.0757 | 0.930 |
| trial count [4th degree] | -0.0021 | -0.0802 – 0.0760 | 0.957 |
| trial count [5th degree] | -0.0729 | -0.1463 – 0.0005 | 0.051 |
| relation length | -0.0030 | -0.0049 – -0.0011 | **0.002** |
| SOI [linear] | 0.0223 | -0.1634 – 0.2081 | 0.813 |
| SOI [quadratic] | -0.1223 | -0.2662 – 0.0216 | 0.095 |
| SOI [cubic] | -0.1197 | -0.2368 – -0.0026 | **0.045** |
| SOI [4th degree] | 0.1204 | 0.0201 – 0.2208 | **0.019** |
| SOI [5th degree] | -0.0405 | -0.1570 – 0.0760 | 0.493 |
| **Random Effects** | | | |
| $\sigma^2$ | 0.0409 | | |
| $\tau_{00\ dyad}$ | 0.0270 | | |
| ICC | 0.3979 | | |
| $N_{dyad}$ | 16 | | |
| Observations | 182 | | |
| Marginal $R^2$ / Conditional $R^2$ | 0.295 / 0.576 | | |

*Table 8: Overview of the LME model of the quantity of motion (estimates, confidence intervals, p-values, random effect and residual variances)*

|  | Wavelet coherence | | |
| --- | --- | --- | --- |
| *Predictors* | *Estimates* | *CI* | *p* |
| (Intercept) | 0.2694 | 0.2489 – 0.2917 | **<0.001** |
| partner realism [2] | 1.0714 | 0.9723 – 1.1806 | 0.161 |
| partner realism [3] | 1.1161 | 1.0040 – 1.2407 | **0.042** |
| musical background [1] | 0.9239 | 0.8529 – 1.0008 | 0.052 |
| trial count [linear] | 0.9130 | 0.8301 – 1.0042 | 0.061 |
| trial count [quadratic] | 1.0969 | 0.9863 – 1.2198 | 0.087 |
| trial count [cubic] | 0.9962 | 0.9057 – 1.0958 | 0.937 |
| trial count [4th degree] | 1.0276 | 0.9282 – 1.1376 | 0.596 |
| trial count [5th degree] | 0.8969 | 0.8124 – 0.9901 | **0.032** |
| training | 1.0483 | 1.0092 – 1.0889 | **0.016** |
| Observations | 88 | | |
| $R^2$ / $R^2$ adjusted | 0.298 / 0.217 | | |

*Table 9: Overview of the LM model of the wavelet coherence of the ternary and binary task (estimates, confidence intervals, p-values, random effect and residual variances)*

|  | Agency | | |
|---|---|---|---|
| *Predictors* | *Odds Ratios* | *CI* | *p* |
| 1\|2 | 0.0035 | 0.0004 – 0.0292 | **<0.001** |
| 2\|3 | 0.0295 | 0.0101 – 0.0865 | **<0.001** |
| 3\|4 | 0.1377 | 0.0586 – 0.3236 | **<0.001** |
| 4\|5 | 0.4910 | 0.2227 – 1.0825 | 0.078 |
| 5\|6 | 3.4340 | 1.5267 – 7.7239 | **0.003** |
| 6\|7 | 64.6045 | 22.3641 – 186.6272 | **<0.001** |
| partner realism [2] | 1.2060 | 0.6060 – 2.4001 | 0.594 |
| partner realism [3] | 1.8815 | 0.8788 – 4.0283 | 0.104 |
| musical background [1] | 2.7497 | 1.5369 – 4.9198 | **0.001** |
| trial count [linear] | 5.8608 | 2.8441 – 12.0773 | **<0.001** |
| trial count [quadratic] | 0.5796 | 0.2730 – 1.2305 | 0.156 |
| trial count [cubic] | 0.9719 | 0.4957 – 1.9054 | 0.934 |
| trial count [4th degree] | 1.1194 | 0.5411 – 2.3159 | 0.761 |
| trial count [5th degree] | 0.3373 | 0.1675 – 0.6795 | **0.002** |
| **Random Effects** | | | |
| $\sigma^2$ | 3.2899 | | |
| $\tau_{00}$ individual | 1.8782 | | |
| ICC | 0.3634 | | |
| N individual | 32 | | |
| Observations | 182 | | |
| Marginal $R^2$ / Conditional $R^2$ | 0.165 / 0.469 | | |

*Table 10: Overview of the CLMM model of the agency scores of the ternary and binary task (estimates, confidence intervals, p-values, random effect and residual variances)*

|  | Shared agency | | |
|---|---|---|---|
| *Predictors* | *Odds Ratios* | *CI* | *p* |
| 1\|2 | 0.0069 | 0.0016 – 0.0292 | **<0.001** |
| 2\|3 | 0.0299 | 0.0100 – 0.0894 | **<0.001** |
| 3\|4 | 0.0630 | 0.0229 – 0.1729 | **<0.001** |
| 4\|5 | 0.2094 | 0.0820 – 0.5350 | **0.001** |
| 5\|6 | 0.9872 | 0.3982 – 2.4476 | 0.978 |
| 6\|7 | 12.2679 | 4.4862 – 33.5476 | **<0.001** |
| partner realism [2] | 3.1164 | 1.5502 – 6.2651 | **0.001** |
| partner realism [3] | 4.0626 | 1.8951 – 8.7092 | **<0.001** |
| trial count [linear] | 4.8265 | 2.3717 – 9.8219 | **<0.001** |
| trial count [quadratic] | 0.6967 | 0.3314 – 1.4648 | 0.340 |
| trial count [cubic] | 0.7204 | 0.3728 – 1.3923 | 0.329 |
| trial count [4th degree] | 1.4078 | 0.6902 – 2.8713 | 0.347 |
| trial count [5th degree] | 0.4793 | 0.2437 – 0.9429 | **0.033** |
| percussion [linear] | 2.7711 | 1.2647 – 6.0716 | **0.011** |
| relation length | 0.9856 | 0.9723 – 0.9991 | **0.036** |
| **Random Effects** | | | |
| $\sigma^2$ | 3.2899 | | |
| $\tau_{00\ individual}$ | 0.8593 | | |
| ICC | 0.2071 | | |
| $N_{individual}$ | 32 | | |
| Observations | 182 | | |
| Marginal $R^2$ / Conditional $R^2$ | 0.228 / 0.388 | | |

*Table 11: Overview of the CLMM model of the shared agency scores of the ternary and binary task (estimates, confidence intervals, p-values, random effect and residual variances)*

POLYRHYTHMIC MUSIC INTERACTION IN EXTENDED REALITY                                                                28|  | SOI | | |
|---|---|---|---|
| *Predictors* | *Odds Ratios* | *CI* | *p* |
| 1\|2 | 0.0309 | 0.0039 – 0.2437 | **0.001** |
| 2\|3 | 0.6623 | 0.1054 – 4.1623 | 0.660 |
| 3\|4 | 2.6565 | 0.4208 – 16.7712 | 0.299 |
| 4\|5 | 6.5190 | 1.0097 – 42.0894 | **0.049** |
| 5\|6 | 31.7711 | 4.6055 – 219.1743 | **<0.001** |
| 6\|7 | 417.1060 | 52.1456 – 3336.3761 | **<0.001** |
| partner realism [2] | 2.3158 | 1.1511 – 4.6591 | **0.019** |
| partner realism [3] | 6.4856 | 2.8325 – 14.8501 | **<0.001** |
| trial count [linear] | 10.4278 | 4.8326 – 22.5010 | **<0.001** |
| trial count [quadratic] | 0.5448 | 0.2548 – 1.1650 | 0.117 |
| trial count [cubic] | 0.6612 | 0.3342 – 1.3082 | 0.235 |
| trial count [4th degree] | 2.7126 | 1.2747 – 5.7725 | **0.010** |
| trial count [5th degree] | 0.4761 | 0.2410 – 0.9408 | **0.033** |
| training | 1.5310 | 1.0229 – 2.2913 | **0.038** |
| **Random Effects** | | | |
| $\sigma^2$ | 3.2899 | | |
| $\tau_{00\ individual}$ | 5.1488 | | |
| ICC | 0.6101 | | |
| $N_{individual}$ | 32 | | |
| Observations | 182 | | |
| Marginal $R^2$ / Conditional $R^2$ | 0.215 / 0.694 | | |

*Table 12: Overview of the CLMM model of the self-other integration scores of the ternary and binary task (estimates, confidence intervals, p-values, random effect and residual variances)*

|  | Flow (absorption) | | |
|---|---|---|---|
| *Predictors* | *Estimates* | *CI* | *p* |
| (Intercept) | 0.6295 | 0.5782 – 0.6808 | **<0.001** |
| musical background [1] | 0.0351 | 0.0113 – 0.0590 | **0.004** |
| SOI y [linear] | 0.0647 | -0.0264 – 0.1559 | 0.163 |
| SOI y [quadratic] | 0.1088 | 0.0333 – 0.1843 | **0.005** |
| SOI y [cubic] | 0.0765 | 0.0146 – 0.1384 | **0.016** |
| SOI y [4th degree] | 0.0147 | -0.0476 – 0.0770 | 0.642 |
| SOI y [5th degree] | 0.0421 | -0.0201 – 0.1044 | 0.183 |
| relation frequency [linear] | -0.0110 | -0.0994 – 0.0774 | 0.806 |
| relation frequency [quadratic] | -0.0713 | -0.1458 – 0.0033 | 0.061 |
| relation frequency [cubic] | 0.0253 | -0.0524 – 0.1030 | 0.521 |
| relation frequency [4th degree] | 0.0457 | -0.0289 – 0.1204 | 0.228 |
| relation length | 0.0012 | 0.0002 – 0.0022 | **0.015** |
| **Random Effects** | | | |
| $\sigma^2$ | 0.0064 | | |
| $\tau_{00 \text{ individual}}$ | 0.0034 | | |
| ICC | 0.3476 | | |
| $N_{\text{individual}}$ | 32 | | |
| Observations | 177 | | |
| Marginal $R^2$ / Conditional $R^2$ | 0.297 / 0.541 | | |

*Table 13: Overview of the CLMM model of the flow (absorption) scores of the ternary and binary task (estimates, confidence intervals, p-values, random effect and residual variances)*